\documentclass[runningheads]{llncs}
\usepackage{graphicx}
\usepackage{enumitem}
\usepackage{pdflscape}

\begin{document}
\title{Debiasing architectural decision-making: a workshop-based training approach}
\titlerunning{Debiasing architectural decision-making}

\author{Klara Borowa\inst{1}\orcidID{0000-0002-7160-5950} \and
        Maria Jarek\inst{1}          \and
        Gabriela Mystkowska\inst{1}  \and
        Weronika Paszko\inst{1} \and
        Andrzej Zalewski\inst{1}\orcidID{0000-0001-5254-4761}}

\authorrunning{K. Borowa et al.}

\institute{Warsaw University of Technology, Institute of Control and Computation Engineering, Warsaw, Poland \\
\email{klara.borowa@pw.edu.pl}\\}

\maketitle              
\begin{abstract}
Cognitive biases distort the process of rational decision-making, including architectural decision-making. So far, no method has been empirically proven to reduce the impact of cognitive biases on architectural decision-making. We conducted an experiment in which 44 master’s degree graduate students took part. Divided into 12 teams, they created two designs -- before and after a debiasing workshop. We recorded this process and analysed how the participants discussed their decisions.
In most cases (10 out of 12 groups), the teams’ reasoning improved after the workshop. Thus, we show that debiasing architectural decision-making is an attainable goal and provide a simple debiasing treatment that could easily be used when training software practitioners.

\end{abstract}
\section{Introduction} \label{introduction}
Cognitive bias is a term that describes an individual’s inability to reason entirely rationally; as such, it prejudices the quality of numerous decisions \cite{van2016decision}. 
Researchers have observed the influence of cognitive biases on day-to-day software development over two decades ago \cite{Stacy1995}. Since then, it was proven that almost all software development activities are affected by cognitive biases to some extent \cite{Mohanani2018}.
Architectural decision-making in particular is not exempt from the influence of cognitive biases \cite{Tang2011}, \cite{van2016decision}. However, research on debiasing architectural decision-making is scarce \cite{Mohanani2018}, with a clear lack of empirically proven debiasing methods that could be used in practice \cite{razavian2019empirical}

In this paper, we endeavour to create an effective debiasing treatment, through expanding on our previous work \cite{borowa2021knowledge}.
The debiasing treatment that we designed consists of an hour-long workshop during which individuals learn about cognitive biases in architectural decision-making (ADM) and take part in three practical exercises. We tested the effectiveness of this debiasing treatment in an experiment, in which 44 master’s level graduate students took part. Our study was aimed at answering the following research question:

\textbf{RQ. Is a training workshop an effective method of reducing the impact of cognitive biases on architectural decision-making? }

Through our study, we show that debiasing ADM is an attainable goal, since in most cases (10 groups out of 12) the debiasing treatment was successful. Our workshop provides a debiasing effect and, because of its simplistic design – it can easily be used to train software practitioners to make more rational decisions.

This paper is organised as follows. In Section \ref{related_work} we describe research related to the subject of our study. Section \ref{method} presents the research method, and in particular: the design of the debiasing workshop, our experiment, the study participants and how we analysed the obtained data. Section \ref{results} contains the results of our experiment. In Section \ref{discussion} we discuss our findings. The threats to validity are explained in Section \ref{threats}. Finally, in Section \ref{conclusion} we provide a conclusion and describe possible future work.

\section{Related work} \label{related_work}
Cognitive biases impact how decisions are made by every human being. In particular, they heavily influence intuitive decisions made under uncertainty \cite{Tversky1974}. 
This effect occurs due to the dual nature of the human mind, which comprises intuitive and rational decision-making subsystems \cite{kahneman2011thinking}. 
Fischoff \cite{fischhoff1982debiasing} describes four levels of debiasing (reducing the effect of biases) treatments: (A) warning about the biases, (B) describing typical biases, (C) providing personalised feedback about the biases, (D) an extended programme of debiasing training.

\noindent \textbf{Cognitive biases influence on architectural decision-making.}
Tang \cite{Tang2011} described how distorted reasoning may impact software design, by providing a set of examples of biased statements that software designers may use during their work \cite{Tang2011}.
As software architecture is actually a set of design decisions \cite{jansen2005software}, it may be heavily affected by architects' biases. This makes reducing the impact of biased decision-making an important endeavour in the area of software architecture  (see also \cite{van2016decision}, \cite{Zalewski2017}).

\noindent \textbf{Debiasing architectural decision-making.}
Although there are various guidelines and practices for improving architectural decision-making \cite{tang2021decision}, \cite{tang2018improving},  there is a severe lack of empirical research on treatments for undesirable behavioural factors in the realm of ADM \cite{razavian2019empirical}. 
There is a small amount of research on debiasing in Software Engineering. 
So far, the existing research has rarely proposed debiasing approaches, and empirical validation of the proposed debiasing methods \cite{Mohanani2018} is even less frequent. 
Notably, Shepperd et al. \cite{Shepperd2018} proposed a successful treatment that improved software effort estimation, through a two- to three-hour-long workshop about cognitive biases.
Our team attempted an empirical validation of an anti-bias treatment for ADM \cite{borowa2021knowledge}, but it turned out not to be successful. 
However, it had several major weaknesses:
\begin{enumerate}
    \item We informed the participants about biases through a simple presentation. This treatment is on the lower levels (A and B) of Fischoff's debiasing scale \cite{fischhoff1982debiasing}.
    In comparison, the successful treatment proposed by Shepperd et al. \cite{Shepperd2018} included a workshop and giving personalised feedback (level C debiasing).
    \item In order to evaluate whether the treatment provided the desired effect, we compared the performance of two groups of students -- one that was shown the presentation, and one that was not. However, this approach does not take into account the teams' individual traits. Those traits may make them more or less susceptible to cognitive biases from the start. It is possible that, when comparing a single team's performance before and after the presentation, the results may have been significantly different.
    \item The sample (2 groups consisting of 5 students) was rather small.
\end{enumerate}
This paper summarises our subsequent research that was aimed at developing a successful debiasing treatment by overcoming the above shortcomings.

\section{Research Method}
\label{method}
The three-hour-long experiment was performed during a meeting on the MS Teams platform.
The experiment plan has been made available online \cite{AdditionalMaterialDebiaisng2022}. 
While planning the experiment, we enhanced most steps from our previous approach \cite{borowa2021knowledge} to both improve the debiasing treatment itself and the validity of the experiment. The basic steps of the  experiment included:
\begin{enumerate}
    \item Preparing the debiasing workshop.
    \item Gathering participants.
    \item A series of three-hour long meetings during which we conducted the experiment, which consisted of three steps:
    \begin{enumerate}
        \item Task 1 -- a 1 hour-long ADM task. 
        \item The debiasing workshop.
        \item Task 2 --  a 1 hour-long ADM task. 
        \end{enumerate}
    \item Analysing the teams’ performance during the first and second tasks.
\end{enumerate}

\subsubsection{Biases}
The debiasing workshop was designed to counter three biases that in previous research turned out to be exceptionally influential on architectural decisions \cite{van2016decision}, \cite{borowa2021influence} and their impact on software engineering overall has already been researched extensively \cite{Mohanani2018} :
\begin{enumerate}
    \item Anchoring -- a biased preference towards initial starting points, ideas, solutions \cite{Tversky1974}.
    \item Confirmation bias -- when the currently desired conclusion leads the individual to search for confirming evidence, or omitting other information \cite{Zalewski2017}.
    \item Optimism bias -- an inclination towards overly optimistic predictions and judgements \cite{Mohanani2018}.
\end{enumerate}

\subsubsection{Architectural decision-making task} \label{task}

Each team of participants performed the task twice: before and after the debiasing workshop. The theme (the problem that was to be solved) was different in each task.
The task was to design an architecture that could be used as a solution to a given theme, and to record the design using the C4 model notation \cite{brown_2018}. 
The task itself was known to the participants before they took part in the experiment, in order to allow them to prepare and learn more about the C4 model. This was not the case for the themes.
All the tasks were supposed to be graded as part of the students’ software architecture course. However, the students were given over a week after the experiment to finish and polish their design. 
During both architectural design tasks, the researchers did not take an active part in the architecting. 

\subsubsection{Debiasing Workshop Design} \label{workshop}
The full workshop plan with instruction for workshop organisers have been made available online \cite{AdditionalMaterialDebiaisng2022}.
The workshop was designed to teach three debiasing techniques:
\begin{itemize}
    \item The anti-anchoring technique: having proposed an architectural solution, the individual that presents it must explicitly list one disadvantage of the solution.
    \item The anti-confirmation bias technique: one team member has to monitor the discussion for unjustified statements that dismiss new information and ideas. Such as ``We already decided that''.
    \item The anti-optimism bias technique: the team must explicitly mention the risks associated with the design decisions.
\end{itemize}

 These specific techniques were proposed previously as a result of our previous work  \cite{borowa2021knowledge} where we analysed, in detail, how each of the three researched biases usually impacted the teams that took part in the study. However, during that study, the effectiveness of these techniques was not validated. For each of these three techniques, the participants had to actively perform a practical exercise.
In this phase of the experiment, the researchers actively facilitated the workshop by encouraging participants to use the debiasing techniques, providing them with examples of the techniques' use, and prompting the participants when they forgot to use the technique that they were supposed to apply.

\subsubsection{Sample} \label{sample}
The participants were recruited from among master’s level graduate students majoring in Computer Science in our Faculty. 
These graduate students in particular, were taking a Software Architecture course. Although participation could be part of their graded project, it was voluntary. There was an alternative, traditional way, to obtain a grade.
At the start of the MS Teams meeting, participants filled a questionnaire that allowed us to obtain basic data about them.
Overall, 61\% of the participants had prior experience in software development, ranging from 0.3 to 3 years. 
The questionnaire and its results, containing detailed information about the participants, is available online \cite{AdditionalMaterialDebiaisng2022}.

\subsubsection{Analysis} \label{analysis}
For the analysis, we used a modified approach of our method from our previous study \cite{borowa2021knowledge}.
In order to analyse the results, we transcribed all recordings of the tasks, during which the participants' created their design. 
In order to inspect how biases impacted architectural decisions, we applied the hypothesis coding method \cite{Saldana2013}. This means that we defined a set of codes to be used to mark relevant segments in the transcript in advance, prior to the analysis. The coding scheme and all specific code counts have been made available online \cite{AdditionalMaterialDebiaisng2022}.

Each transcript was first coded by two researchers separately. Then, all of the codes were negotiated \cite{Garrison2006} until the coders reached a consensus on each code. 
Additionally, no transcript was coded by the same researcher that conducted the particular meeting with the participants. Furthermore, we summarised the overall number of codes only twice, after coding 6 and 12 transcripts, to avoid a situation where we would unconsciously chase after a desired number of biased or non-biased arguments in a particular transcript. 

Having coded the transcripts, we compared how many biased and non-biased statements/decisions were present before and after the workshop. 
We defined: (a) a biased decision as one impacted by more biased statements than rational arguments, (b) a non-biased decision as one impacted by more rational arguments than biased statements, (c) neutral decisions as ones impacted by an equal amount of biased and non-biased statements. 
We also counted the amount of bias influences and the usage of the debiasing techniques
during the tasks.

\section{Results} \label{results}
Through the analysis process we uncovered the specifics about arguments, decisions, bias occurrences and the use of debiasing techniques 
in the teams' Task 1 and Task 2 transcripts. All p-values mentioned in this section were calculated using the non-parametric Wilcoxon Signed Rank Test. Through this test, we evaluated whether the changes in specific measured values were statistically different (when the p-value was less than 0.05).
All specific numbers for code counts for each team are available online \cite{AdditionalMaterialDebiaisng2022}.

\subsubsection{Arguments.} \label{arguments}
We classified these arguments as either biased (i.e. affected by one or more of the researched biases) or non-biased. Overall, we found 1470 arguments and 487 counterarguments. 54\% of the statements before the workshop were biased, compared to 36\% after. In general, the percentage of biased arguments decreased after the workshop in the cases of all teams except one.

The increased number of non-biased arguments (p-value = 0.0024) and non-biased counterarguments (p-value = 0.0005), and the decrease of the percentage of biased statements (p-value = 0.002) were significant.  However, the changes in the number of biased arguments (p-value = 0.1973) and counterarguments (p-value = 0.8052) can not be considered significant. 

\subsubsection{Decisions.}\label{decisions}
Overall we found 641 decisions - 266 biased, 281 non-biased and 94 neutral. 52\% of decisions before the workshop were biased, compared to 31\% after. 
Only one had a larger percentage of biased decisions after the workshop. In the case of all the other teams, the percentage of biased decisions decreased.
The increase in the number of non-biased decisions (p-value = 0.0024) and the decreased percentage of biased decisions (p-value = 0.0020) were significant. However, the change in the number of biased decisions (p-value = 0.0732) can not be considered significant.

\subsubsection{Cognitive biases.}\label{biases}
Overall, we found 1110 bias occurrences - 558 before and 552 after the workshop.
The sum of these counts is different from the number of arguments since: (a) it was possible for various biases to influence one argument, (b) some biased statements were not connected to any architectural decision.

There was no significant change in the overall number of biases between Task 1 and Task 2 (p-value = 0.8647).
This means that the debiasing effect (the smaller percentage of biased decisions and arguments) was not achieved by decreasing the number of bias occurrences during the tasks.
In fact, the effectiveness of the debiasing treatment comes from increasing the number of non-biased arguments.

\subsubsection{Debiasing techniques.} \label{debiasing}
We compared the amounts of technique uses before and after the workshop, since it was possible for participants to spontaneously use a specific technique during Task 1. We identified 133 uses of the proposed techniques - 26 techniques before and 107 after the workshop.

The number of uses of the practices increased significantly during Task 2 (p-value = 0.0005). However, three teams did not increase their use of the anti-bias techniques substantially after the workshop. Despite this, these teams’ percentage of biased arguments and decisions decreased during Task 2. Additionally, two teams, despite using a higher number of debiasing techniques during Task 2, had more biased decisions and more biased arguments during Task 2. 

Overall, the anti-optimism technique was used most often (15 before and 57 after workshop), while the anti-anchoring technique was used less often (3 before and 30 after workshop), with the anti-confirmation bias technique rarely being used at all (8 before and 20 after workshop). This may be because listing risks came most naturally,
while the other two techniques may require much more effort to be used correctly.

\section{Discussion}  
\label{discussion}
Our results show that the debiasing treatment through the debiasing workshop we designed was successful, both improving the quality of argumentation and design decisions.
However, there are some particularities worth discussing in detail, which may help to significantly improve our approach in the future.

Firstly, our approach did not significantly decrease the number of cognitive bias occurrences, biased arguments and biased decisions that impacted our participants (see Section \ref{results}). Instead, we managed to improve the number of rational arguments and decisions present in the teams’ discussions, through which the percentage of biased arguments and decisions decreased. This means that, while  it may not be possible to completely get rid of cognitive biases, other ways of rationalising decision-making are possible and could be pursued.

Secondly, the team whose decisions improved the most was the one that had the worst result in Task 1. Furthermore, the team that improved the least was the one with the best result in Task 1. 
This may mean that, while our treatments successfully improve the performance of initially biased individuals, it may not be as impactful in the case of individuals that were initially less impacted by biases. Since our participants were students with up to three years of professional experience, we do not know yet how different the debiasing effect would be on experienced practitioners (who may be initially less impacted by biases).

Finally, the failure of two teams led us to explore the transcripts in detail. After that, we noticed that their performance dropped at one point, when the participants simply became tired (which they expressed verbally).
This is in line with Kahneman’s \cite{kahneman2011thinking} explanation for the existence of two systems – using System 2 is physically exhausting and no human can use it indefinitely. Thus, time for rest may be a crucial factor to bear in mind while attempting debiasing.

\section{Threats to Validity}
\label{threats}

\textbf{Conclusion validity}: 
We used non-parametric tests to examine whether the observed changes in the measured values were significant.

\textbf{Internal validity}: 
We put significant care into designing the experiment and setting the environment so that no factors other than the workshop influenced the students. 
The teams did not know the themes for their tasks before the study, and did not have more than 10 minutes of time to interact with the environment outside during the experiment. 
Finally, to decrease the chances of the researchers distorting the results during the analysis, we used negotiated coding \cite{Garrison2006} and calculated the results (code counts) for the transcripts only twice.

\textbf{Construct validity}: 
Our method also improves on the one used in our previous study \cite{borowa2021knowledge} by taking into account not only arguments and biases but also decisions. This factor is crucial since it is possible that, while the overall argumentation may improve, a team can lack regularity when using rational arguments, thus still making numerous biased decisions nonetheless.

\textbf{External validity}: 
Our study’s weakness is that our participants were all students. While most of them had professional experience, it was limited.

\section{Conclusion and Future Work} 
\label{conclusion}
In this paper we explored whether debiasing through a training workshop is an effective method of reducing the impact of cognitive biases on ADM (RQ).
We designed such a workshop and examined its effectiveness (Section \ref{method}). The results show that the debiasing treatment is effective (Section \ref{results}), although it does not completely eliminate the impact of the biases (Section \ref{discussion}).

Through this work, we show that designing a successful debiasing treatment for cognitive biases in ADM is possible, and propose an effective treatment that can become a foundation for future research.  
Future research can focus on: testing different debiasing techniques, debiasing that takes into account other cognitive biases, exploring the workshop's effectiveness in debiasing experienced practitioners.

Practitioners can use the presented debiasing workshop for training purposes.

\bibliographystyle{splncs04}
\bibliography{references}
\end{document}